\documentclass{prop2015}
\usepackage[english]{babel}

\title{Solutions in Exceptional Field Theory}

\author[F. J. Rudolph]{Felix J. Rudolph\inst{1,}\footnote{E-mail:~\textsf{f.j.rudolph@qmul.ac.uk}}}
\address[1]{Queen Mary University of London, Centre for Research in String Theory, School of Physics, Mile End Road, London, E1 4NS, England}
\shortauthors{F. J. Rudolph}

\begin{abstract}
Exceptional Field Theory employs an extended spacetime to make supergravity fully covariant under the U-duality groups of M-theory. This allows for the wave and monopole solutions to be combined into a single solution which obeys a twisted self-duality relation. All fundamental, solitonic and Dirichlet branes of ten- and eleven-dimensonal supergravity may be extracted from this single solution in Exceptional Field Theory.
\end{abstract}
\shortabstract

\begin{document}
\maketitle

\section{Exceptional Field Theory}

The primary idea behind Exceptional Field Theory \cite{Hohm:2013pua} is to make the exceptional symmetries of eleven-dimensional supergravity manifest. The appearence of the exceptional groups in dimensionally reduced supergravity theories was first discussed in \cite{Cremmer:1998}. Exceptional Field Theory (EFT) is a similar construction as Doubel Field Theory (DFT, see \cite{Aldazabal:2013sca,Berman:2013eva,Hohm:2013bwa} for reviews and references) which makes T-duality manifest. 

In EFT one first performes a decomposition of eleven-dimensional supergravity -- but with no reduction or truncation -- into an $(11-d) \times d$ split. Then one supplements the $d$ so-called ``internal'' directions with additional coordinates to linearly realize the exceptional symmetries. That is one extends the eleven dimensions of supergravity to
\begin{equation}
M^{11} = M^{11-d} \times M^d \longrightarrow M^{11-d} \times M^{\dim E_d}
\end{equation}
where $\dim E_d$ is the dimension of the relevant representation of the exceptional group $E_d$ and $M^{\dim E_d}$ is a coset manifold that comes equipped with the coset metric of $E_d/H$ (where $H$ is the maximally compact subgroup of $E_d$). 

The U-duality groups are related to the embedding of the eleven dimensions in the extended space. The combination of $p$-form gauge transformations and diffeomorphism give rise to a continuous local $E_d$ symmetry. This however is not U-duality which is a global discrete symmetry that only occurs in the presence of isometries. (See \cite{Berman:2014jba} for the equivalent discussion of T-duality in DFT). 

Crucially however there is also a \emph{physical section condition} that provides a constraint in EFT that restricts the coordinate dependence of the fields to a subset of the dimensions and thus there naturally appears a physical submanifold which is identified as the usual spacetime. When there are no isometries present this section condition constraint produces a canonical choice of how spacetime is embedded in the extended space. However, in the presence of isometries there is an ambiguity in how one identifies the submanifold in the extended space. This ambiguity is essentially the origin of U-duality with different choices of spacetime associated to U-duality related descriptions. (This is discussed in detail for the case of DFT in \cite{Berman:2014jsa} and for EFT in \cite{Berman:2014hna}).

To be more specific, we will pick $d=7$ for the following construction. First the basics of the $E_7$ EFT are summarized. Then a specific solution in that theory is presented and its relation to supergravity solutions is discussed. This EFT solution provides a good example of how the isometries and ambiguity of picking the physical spacetime lead to U-duality.

\section{The $E_7$ EFT}

If eleven-dimensional supergravity is split into $4+7$ dimensions, the U-duality group is $E_7$. It can be made into a manifest symmetry by extending the seven internal dimentions to a 56-dimensional space (the $\mathbf{56}$ is the fundamental representation of $E_7$). The field content of this $4+56$-dimensional EFT \cite{Hohm:2013uia} is
\begin{equation}
\Big\{ g_{\mu\nu}, {\mathcal{A}_\mu}^M, \mathcal{M}_{MN}, \mathcal{B}_{\mu\nu\, M} \mathcal{B}_{\mu\nu\, \alpha},  \Big\} \, .
\label{eq:fields}
\end{equation}
Here $g_{\mu\nu}$ is the spacetime metric of the external space ($\mu=1,\dots,4$) and $\mathcal{M}_{MN}$ is the generalized metric of the extended internal space ($M=1,\dots,56$) which parametrizes the coset $E_7/SO(56)$. The generalized tangent space of the extended space is isomorphic to a sum of tensor bundles
\begin{equation}
TM \oplus \Lambda^2 T^*M \oplus \Lambda^5 T^*M \oplus (T^*M\otimes\Lambda^7 T^*M)
\label{eq:tangentspace}
\end{equation}
where $M=M^7$ is the seven-space. The terms in the sum correspond to brane charges: momentum, membrane, fivebrane and KK-monopole charge. These are the conjugate variables to the 56 coordinates of the extended space which can be seen as brane \emph{wrapping} coordinates in analogy to the string \emph{winding} coordinates of DFT.

The ``cross-term'' ${\mathcal{A}_\mu}^M$ is a set of 56 vector fields with field strength ${\mathcal{F}_{\mu\nu}}^M$ which obeys a twisted self-duality relation
\begin{equation}
{\mathcal{F}_{\mu\nu}}^M = \frac{1}{2}\sqrt{g}\epsilon_{\mu\nu\rho\sigma}
							\Omega^{MK}\mathcal{M}_{KN}\mathcal{F}^{\mu\nu\, N} 
\label{eq:duality}
\end{equation}
where $g$ is the determinant of the external metric and $\Omega$ is the invariant symplectic form of $Sp(56)\supset E_7$.
The two auxiliary two-forms $\mathcal{B}_{\mu\nu}$ are required in the tensor hierarchy of the supergravity. They carry a fundamental index $M$ and an adjoint index $\alpha$ respectively ($\alpha=1,\dots,133$). The action integral and a detailed discussion of these fields can be found in \cite{Hohm:2013uia}.

\section{The Self-dual Solution}

In \cite{Berkeley:2014nza} it was shown how a null wave carrying momentum in one of the novel dimensions of DFT or EFT corresponds to a fundamental string or membrane in supergravity whose  mass and charge is given by the momentum of the wave. Similarly it was shown in \cite{Berman:2014jsa} how a monopole structure (Hopf fibration) in the extended space gives the solitonic fivebrane in supergravity. 

These two approaches are combined into a single self-dual solution of EFT with U-duality group $E_7$ \cite{Berman:2014hna}. The solution has both wave-like and monopole-like aspects. It is a null wave carrying momentum which correspondes to electric charge \`a la Kaluza-Klein. It is also equipped with a monopole structure whose non-trivial first Chern class provides the magnetic charge. 

The fields \eqref{eq:fields} of the solution are given in terms of a harmonic function which only depends on the external coordiantes and is smeared over the extended internal directions. The vector fields ${\mathcal{A}_\mu}^M$ have only two non-zero components which are related by the duality \eqref{eq:duality} and reflect the ``electric'' and ``magnetic'' -- i.e. wave and monopole -- aspects of the solution. The solution thus has a characteristic direction (given by the electric component) whose radius is associtated with the quantization of the relevant charge.

\section{Relation to Supergravity Solutions}

From a conventional supergravity point of view in ten or eleven dimensions, the EFT solution corresponds to all 1/2 BPS branes depending on how the physical spacetime is picked out in the extended space. The supergravity fields such as the metric, the 3-form and the 6-form potentials in eleven dimensions (or the dilaton and the NSNS and RR form fields in ten dimensions) can all be recovered from the EFT fields. 

The solution only depends on the external coordinates with isometries in the internal space and therefore clearly obeys the physical section condition. But there is an ambiguity of how to pick the six or seven dimensions out of the 56 to form the physical spacetime. It is precisely this ambiguity which leads to the different supergravity solutions which are all related by U-duality. 

To be more concrete, if for example the characteristic direction of the EFT solution is along one of the membrane wrapping directions in \eqref{eq:tangentspace}, it appears as the membrane in the supergravity picture. Similarly for the other possible directions. It is also possible to have the characteristic direction of the solution in a superposition of, say, a membrane and fivebrane direction which will lead to a bound state solution.

To conclude, in the framework of Exceptional Field Theory there is a single, self-dual solution which corresponds to all 1/2 BPS branes in supergravity which are related by U-duality.

\end{document}